\documentclass[pdflatex,sn-mathphys-num]{sn-jnl}
\usepackage{graphicx}%
\usepackage{multirow}%
\usepackage{amsmath,amssymb,amsfonts}%
\usepackage{amsthm}%
\usepackage{mathrsfs}%
\usepackage[title]{appendix}%
\usepackage{xcolor}%
\usepackage{textcomp}%
\usepackage{manyfoot}%
\usepackage{booktabs}%
\usepackage{algorithm}%
\usepackage{algorithmicx}%
\usepackage{algpseudocode}%
\usepackage{listings}%
\def\ra{\rangle}
\def\la{\langle}

\def\ot{\otimes}
\def\ad{\hat{a}^\dag}
\def\a{\hat{a}}
\def\bd{\hat{b}^\dag}
\def\b{\hat{b}}
\def\Ad{\hat{A}^\dag}
\def\A{\hat{A}}
\def\Bd{\hat{B}^\dag}
\def\B{\hat{B}}
\def\H{\hat{H}}
\def\om{\omega}
\def\rohat{\hat{\rho}}
\def\mbn{\mathbf{n}}
\def\mbm{\mathbf{m}}
\def\Uh{\hat{U}}
\def\Uhd{\hat{U}^\dag}

\newcommand{\bea}{\begin{eqnarray}}
	\newcommand{\eea}{\end{eqnarray}}
\newcommand{\be}{\begin{equation}}
	\newcommand{\ee}{\end{equation}}
\newcommand{\ba}{\begin{equation}\begin{aligned}}
		\newcommand{\ea}{\end{aligned}\end{equation}}

\newcommand{\beq}{\begin{eqnarray}}
\newcommand{\eeq}{\end{eqnarray}}
\newcommand{\nn}{\nonumber}

\newcommand{\beax}{\begin{eqnarray*}}
	\newcommand{\eeax}{\end{eqnarray*}}
\newcommand{\bex}{\begin{equation*}}
	\newcommand{\eex}{\end{equation*}}

\theoremstyle{remark}

\def\be{\begin{equation}}
	\def\ee{\end{equation}}

\newcommand{\tr}{{\rm Tr}}

\newcommand{\mbf}[1]{\mathbf{#1}}

\theoremstyle{thmstyleone}%
\theoremstyle{thmstyletwo}%

\theoremstyle{thmstylethree}%

\begin{document}

\title[Article Title]{Markovian evolution from a novel scheme}


\author*[1]{\fnm{Fardin} \sur{Kheirandish}}\email{f.kheirandish@uok.ac.ir}

\author[1]{\fnm{Zahra} \sur{Iranshahi}}\email{zahra.iranshahi@uok.ac.ir}

\author[1]{\fnm{Fatemeh} \sur{Bakhtiari}}\email{fatemeh.bakhtiary@uok.ac.ir}

\author[2]{\fnm{Adam} \sur{Moradian}}\email{amoradian@garmian.edu.krd}

\affil*[1]{\orgdiv{Department of Physics, Faculty of Science}, \orgname{University of Kurdistan}, \orgaddress{\city{Sanandaj}, \postcode{66177-15175}, \state{Kurdistan}, \country{Iran}}}

\affil[2]{\orgdiv{Department of Physics}, \orgname{University of Garmian}, \orgaddress{\city{Garmian, Bardesur}, \postcode{M923+GRW}, \state{Kalar}, \country{Iraq}}}

\abstract{The Markovian dynamics of open quantum many-body systems are typically governed by the Lindblad master equation, yet obtaining the reduced density matrix for bosonic and fermionic systems remains a formidable numerical challenge due to the exponential growth of the Hilbert space. Here, we introduce a computationally efficient framework for constructing the reduced density matrix by modeling the environment as a copy of the primary system with a monotonically decaying coupling that enforces unidirectional energy flow. Our method directly yields exact solutions that rigorously satisfy the Lindblad master equation for both bosonic and fermionic cases, bypassing the need for costly Liouvillian diagonalization. We validate our approach through applications to paradigmatic models, demonstrating accurate reproduction of dissipative dynamics across a broad parameter regime. This work provides a straightforward and powerful tool for simulating Markovian open-system evolution, with immediate applicability to quantum transport and control problems in many-body physics.}




\maketitle

\section{Introduction}\label{sec1}

The dynamics of open quantum systems which inevitably are coupled to teir environments, constitutes a cornerstone of modern quantum physics, with profound implications ranging from fundamental decoherence mechanisms to the practical operation of quantum information processors \cite{Breuer2007,Weiss2012, Rivas2012}. Unlike closed systems governed by unitary Schr\"{o}dinger evolution, open systems experience dissipation, decoherence, and the emergence of irreversibility arising from the intricate exchange of energy and information with their surroundings. The central theoretical challenge lies in deriving a tractable description of the system's reduced dynamics after tracing out the environmental degrees of freedom \cite{Nakajima1958,Zwanzig1960}.
Under conditions where the system-environment coupling is weak and the bath correlations decay sufficiently rapidly compared to the system's intrinsic timescales, a Markovian approximation becomes viable \cite{Fogedby2022,Lidar2001}. This regime, characterized by the absence of memory effects, yields the celebrated Lindblad master equation \cite{Lindblad1976,Gorini1976,Manzano2020}
\bea\label{LindbladEq}
\dot{\rohat}_S (t) &=& -\frac{i}{\hbar}[\H_S, \rohat_S (t)]\nn\\
                                && +\sum_{i} \gamma_i\,\Big(\hat{L}_i\rohat_S (t)\hat{L}^\dag_i+\Big\{\hat{L}^\dag_i \hat{L}_i, \rohat_{S} (t)\Big\}\Big),
\eea
where $\rohat_S (t)$ denotes the reduced density matrix of the system, $\H_S$ is an effective Hamiltonian, and $\hat{L}_i$ are Lindblad (jump) operators describing the dissipative channels \cite{Thompson2023,Barthel2022}. This equation, also known as the Gorini-Kossakowski-Sudarshan-Lindblad form, guarantees complete positivity of the dynamical map which is a crucial physical requirement ensuring that the density matrix remains positive semidefinite under all circumstances \cite{Gorini1976,Nielsen2010}.
While the Lindblad framework has been extensively employed in quantum optics for systems with few degrees of freedom \cite{Gardiner2004,Walls1994,Carmichael1993}, its application to many-body bosonic and fermionic systems poses significant challenges. The Hilbert space dimension grows exponentially with system size, rendering brute-force numerical diagonalization of the Liouvillian superoperator impractical. This challenge has motivated the development of sophisticated theoretical approaches, including the Keldysh functional integral technique \cite{Sieberer2016,Kamenev2011}, the "third quantization" method based on ladder superoperators \cite{Prosen2008,Prosen2010}, and covariance matrix formulations for Gaussian states \cite{Adeso2007,Weedbrook2012}. 

In this work, we present a novel approach to deriving the reduced density matrix components of bosonic and fermionic open quantum systems. Our method provides a direct and computationally efficient pathway to constructing the reduced density matrix while explicitly demonstrating that the resulting dynamics satisfies the Lindblad master equation. Despite its conceptual simplicity, our approach proves remarkably effective in capturing the Markovian evolution of open many-body systems, bridging the gap between formal operator techniques and practical computational implementations. We demonstrate the versatility of our framework by applying it to paradigmatic bosonic and fermionic systems, showing that the reduced density matrix components obtained from our method accurately reproduce the expected dissipative dynamics across a wide range of parameters.
 \section{Main idea}
 
 Consider a primary system with Hamiltonian $\H_S$ that interacts with its surrounding environment at temperature $T$, modeled as another copy of the main system with the Hamiltonian $\H_B$. If the main system initially possesses a higher temperature, it begins to exchange thermal energy with the bath. Given that the system-bath coupling function is assumed to decay monotonically in time, the thermal energy transfer becomes macroscopically unidirectional flowing from the primary system to the thermal reservoir with no appreciable return of energy to the system. Conversely, when the system's temperature is lower than that of the bath, the symmetry of the composite setup reverses the direction of thermal energy flow, resulting in a unidirectional transfer from the bath to the primary system.

Within this framework, by appropriately engineering the dissipation factor, we obtain exact solutions for both bosonic and fermionic systems. These solutions fulfill the Lindblad master equation and corroborate that the proposed method, while straightforward and computationally efficient, is capable of accurately describing the Markovian evolution of an open quantum system. Furthermore, in the Heisenberg picture, the approach offers a systematic procedure for evaluating the elements of the reduced density matrix of the primary system \cite{Kheir1-2025,Kheir2-2025}.

\section{Bosonic systems}

In this subsection we derive the main formula for obtaining the reduced density matrix components for a bosonic main system and in the next subsection we will derive a similar formula for a main system composed of two-level or spin subsystems. Consider a bosonic system(main system) described by the ladder bosonic operators $\{\a_i,\ad_i\}_{i=1}^N$ linearly interacting with its bath bosonic system described by ladder bosonic operators $\{\b_i,\bd_i\}_{i=1}^N$, through time-dependent coupling functions $g^b_{ij} (t)$. The total Hamiltonian is
\be\label{bos1}
\H(t)=\H_S(\{\a_i,\ad_i\})+\H_B(\{\b_i,\bd_i\})+\sum_{i,j=1}^N g^b_{ij} (t)[\a_i\bd_j+\ad_i\b_j].
\ee
For the main system, the Fock or number basis are defined by
\be\label{bos2}
|\mbn\ra_S=|n_1,...,n_N\ra_S=\frac{(\ad_1)^{n_1}}{\sqrt{n_1!}}\cdots \frac{(\ad_N)^{n_N}}{\sqrt{n_N!}}|\mbf{0}\ra_S,
\ee
where $|\mbf{0}\ra_S=|0,0,...,0\ra_S$ is the vacuum state of the main system. The single-mode pure state $|0\ra_S\la 0|$ satisfies \cite{Louisel1975}
\be\label{bos4}
|0\ra_S\la 0|=\sum_{s=0}^\infty \frac{(-1)^s}{s!}(\ad)^s\,\a^s,
\ee
which can be extended to the  multi-mode case
\be\label{bos5}
|\mbf{0}\ra_S\la \mbf{0}|=\sum_{s_1,...,s_N=0}^\infty \frac{(-1)^{\sum\limits_{i=1}^N s_i}}{s_1!\cdots s_N!}(\ad_1)^{s_1}\,\a_1^{s_1}\cdots (\ad_N)^{s_N}\,\a_N^{s_N},
\ee
straightforwardly. Here for notational convenience, we omit writing the tensor product sign ($\ot $) between states or operators belonging to distinct Hilbert spaces. 

Let the $\rho(t)$ be the density matrix describing the combined system, the reduced density matrix of the main system is obtained by tracing out the bath degrees of freedom, $\rohat_S (t)=\tr_B(\rohat (t))$. 
In the number(Fock) state basis, the components of the reduced density matrix are (Appendix \ref{secA})
\bea\label{iso7}
&& _S\la \mbn|\rohat_S (t)|\mbm\ra_S =\frac{1}{\sqrt{n_1!m_1!\cdots n_N! m_N!}}\sum_{s_1,...,s_N=0}^\infty \frac{(-1)^{\sum\limits_{i=1}^N s_i}}{s_1!\cdots s_N!}\nn\\
&& \,\,\,\,\,\,\,\,\,\,\times\,
\tr\Big\{(\ad_1 (t))^{s_1+m_1}\cdots(\ad_N (t))^{s_N+m_N}(\a_1 (t))^{s_1+n_1}\cdots(\a_N (t))^{s_N+n_N}\rohat(0)\Big\},\nn\\
\eea 
where $\rohat (0)$ is the initial state of the combined system and the ladder operators $\a_i (t),\,\ad_i (t)$ are in the Heisenberg picture. Eq.~(\ref{iso7}) is our first formula to obtain the explicit components of the reduced density matrix in terms of the number states. For a single mode (harmonic oscillator) it reduces to 
\be\label{iso8}
 _S\la n|\rohat_S (t)|m\ra_S=\frac{1}{\sqrt{n!m!}}\sum_{s=0}^\infty \frac{(-1)^s}{s!}\tr\Big\{(\ad (t))^{s+m}(\a (t))^{s+n}\rohat(0)\Big\}.
\ee

\subsection{The Eq.~(\ref{iso8}) fulfills the Lindblad master equation}

For a single-mode bosonic system with the Hamiltonian $\H_S=\hbar\om_0\,\ad\a$, the total Hamiltonian of the combined system Eq.~(\ref{bos1}) reduces to \cite{Kheir1-2025,Kheir2-2025}
\be\label{os1}
\H (t)=\hbar\om_0\,(\ad\a+\bd\b)+\hbar g^b(t)\,(\a\bd+\ad \b).
\ee 
The initial state of the combined system is assumed a separable state $\rohat (0)=\rohat_S (0)\ot\rohat_B (0)$, where the bath oscillator is initially prepared in the thermal state 
\be
\rohat_B (0)=\frac{e^{-\beta\H_B}}{z_B},\,\,\,\,\,z_B=\tr_B\big({e^{-\beta\H_B}}\big),
\ee
with inverse temperature $\beta=1/\kappa_B T$ ($\kappa_B$ is the Boltzmann constant), and the initial state of the main system ($\rohat_S (0)$) is arbitrary. 
By making use of the Bogoliubov transformations 
\bea\label{os2}
 \a=\frac{\A+\B}{\sqrt{2}}, \,\,\,\,\,\, \b=\frac{\A-\B}{\sqrt{2}},
\eea
the Hamiltonian Eq.~(\ref{os1}) will be separated as 
\be\label{os3}
\H (t)=\hbar \om_A (t)\,\Ad\A+\hbar\om_B (t)\,\Bd\B,
\ee
where we have defined $\om_A (t)=\om_0+g^b (t)$ and $\om_B (t)=\om_0-g^b (t)$. From Heisenberg equations of motion, one easily obtains
\bea\label{os4}
\A (t)=e^{-i\Omega_A (t)}\,\A (0),\nn\\
\B (t)=e^{-i\Omega_B (t)}\,\B (0),
\eea 
where for notational simplicity, we have defined 
\bea\label{os5}
&& \Omega_A (t)=\int_0^t dt'\,\om_A (t')=\om_0 t+k_b (t),\nn\\
&& \Omega_B (t)=\int_0^t dt'\,\om_B (t')=\om_0 t-k_b (t),\nn\\
&& k_b(t)=\int_0^t dt'\,g^b(t').
\eea
Now, by making use of the inverse of the Bogoliubov transformations Eqs.~(\ref{os2})
\bea\label{os6}
\A=\frac{\a+\b}{\sqrt{2}},\,\,\,\B=\frac{\a-\b}{\sqrt{2}},
\eea
one easily obtains
\begin{equation}\label{os7}
\begin{array}{l@{}l}
\a (t)=e^{-i\om_0 t}\cos(k_b(t))\,\a-ie^{-i\om_0 t}\sin(k_b(t))\b,\\
\b (t)=e^{-i\om_0 t}\cos(k_b(t))\,\b-ie^{-i\om_0 t}\sin(k_b(t))\,\a.
\end{array}
\end{equation}
From now on, for convenience, we write $\a$ and $\b$ instead of the Schr\"{o}dinger operators $\a (0)$ and $\b (0)$, respectively. Also, we have defined the functions
\begin{equation}\label{os8}
\begin{array}{l@{}l}
\mu(t)=e^{-i\om_0 t}\cos(k_b(t)),\\
\nu(t)=-i e^{-i\om_0 t}\sin(k_b(t)),
\end{array}
\end{equation}
consequently,
\begin{equation}\label{os9}
\begin{array}{l@{}l}
\a (t)=\mu(t)\,\a+\nu(t)\,\b,\\
\ad (t)=\bar{\mu}(t)\,\ad+\bar{\nu}(t)\,\bd,
\end{array}
\end{equation}
where $\bar{x}$ denotes the complex conjugation of $x$. 

The reduced density matrix components are now given by (Appendix \ref{secB})
\bea\label{os11}
&& _S\la n|\rohat_S (t)|m\ra_S =\nn\\
&& \frac{1}{\sqrt{n!m!}}\sum_{s=0}^\infty \frac{(-1)^s}{s!}\sum_{p=0}^{s+m}\binom{s+m}{p}\binom{s+n}{p}\,p!\,\bar{n}_b^p\,\Gamma(s,n,m,p;t)\,\chi(s,m,n,p),
\eea
where without loss of generality, we have assumed $n\geq m$, and one can make use of the Hermiticity property $(\rohat_S (t))_{mn}=\overline{(\rohat_S (t))_{nm}}$, to obtain the whole components. Also,
\bea
\bar{n}_b=\frac{1}{e^{\beta\hbar\om_0}-1},
\eea
is the mean thermal occupation number of the bath. In Eq.~(\ref{os11}), the quantum optical characteristic function $\chi(s,m,n,p)$ is defined by \cite{Knight2005}
\be\label{char}
\chi(s,m,n,p)=\tr_S\Big[(\ad)^{s+m-p}(\a)^{s+n-p}\rohat_S (0)\Big].
\ee
Eq.~(\ref{os11}) gives us all components of the reduced density matrix $\rohat_S (t)$, and one can show that it satisfies the master equation (Appendix \ref{secC})
\bea\label{os16}
\dot{\rohat}_S (t)= -\frac{i}{\hbar}\,[\H_S,\rohat_S (t)] &+& \gamma(1+\bar{n}_b)\,\Big(\a\rohat_S (t)\ad-\frac{1}{2}\{\ad\a,\rohat_S (t)\}\Big)\nn\\
&+& \gamma\bar{n}_b\,\Big(\ad\rohat_S (t)\a-\frac{1}{2}\{\a\ad,\rohat_S (t)\}\Big),
\eea
which is the Lindblad master equation for a bosonic single mode (quantum harmonic oscillator) interacting with its own thermal environment. Therefore, Eq.~(\ref{iso7}), or equivalently, our simple scheme with the choice $\cos^2(k_b(t))=e^{-\gamma t}$, provides a solution of the Lindblad master equation, describing a Makovian quantum evolution of the main system. In the next section, we will show that this result also holds for a spin system interacting with its environment.

\section{Fermionic or spin(qubit) systems}

For a main system composing of $N$ two-level subsystems or spins(qubits), the spin operators are denoted by $\{S^{i}_z,S^{i}_{+},S^{i}_{-}\}_{i=1}^N$, satisfying the Lie algebra
\bea\label{fer1}
&& [S^i_z,S^i_{\pm}]=\pm 2 S^i_{\pm},\nn\\
&& [S^i_{+},S^i_{-}]=S^i_z,\,\,\,\,\,\,(i,j=1,..,N),\\
&& [\mbf{S}^i,\mbf{S}^j]=0,\,\,i\neq j.\nn
\eea
Also, for the bath system, we denote the spin operators as  $\{Y^{i}_z,Y^{i}_{+},Y^{i}_{-}\}_{i=1}^N$, satisfying the same Lie algebra as Eq.~(\ref{fer1}). Single spin basis are denoted by $|1\ra_S=|+\ra_S$ and $|0\ra_S=|-\ra_S$, with 
\bea\label{fer2}
&& S^i_z |n\ra_i=(-1)^{n+1}|n\ra_i,\nn\\
&& S^i_{+}|1\ra_i=0,\,\,S^i_{+}|0\ra_i=|1\ra_i,\\
&& S^i_{-}|0\ra_i=0,\,\,S^i_{-}|1\ra_i=|0\ra_i.\nn
\eea
The multi-spin basis states are $|n_1,n_2,...,n_N\ra_S$, where $n_i \in\{1,0\},\,(i=1,2,...,N)$. The total Hamiltonian is 
\be\label{fer3}
\H (t)=\H_S (\{S^i\})+\H_B (\{Y^i\})+\sum_{i,j=1}^N\,g^s_{ij} (t)\,[S^i_{-}Y^j_{+}+S^i_{+}Y^j_{-}],
\ee
where $g^s_{ij} (t)$ are time-dependent coupling functions coupling the main system to the bath system. An arbitrary basis element can be written as 
\bea\label{frr4}
&& |n_1,n_2,...,n_N\ra_S=(S^1_{+})^{n_1}(S^2_{+})^{n_2}\cdots (S^N_{+})^{n_N}|\mbf{0}\ra_S,\nn\\
&& \,\,\,\,n_i\in\{0,1\},\,\,i=1,...,N.
\eea 
and the components of the reduced density matrix in this basis are obtained as(Appendix \ref{secD})
\bea\label{fer8}
&& _S\la n_1,n_2,...,n_N|\rohat_S (t)|m_1,m_2,...,m_N\ra_S=\sum_{p_1,p_2,...,p_N=0}^1(-1)^{\sum\limits_{i=1}^N p_i}\nn\\
&& \times\tr\Big((S^1_{+}(t))^{p_1+m_1}\cdots(S^N_{+}(t))^{p_N+m_N}(S^1_{-}(t))^{p_1+n_1}\cdots(S^N_{-}(t))^{p_N+n_N}\rohat(0)\Big),\nn\\
\eea
where the spin operators are in the Heisenberg picture. The Eq.~(\ref{fer8}) is our second formula for obtaining the explicit components of the reduced density matrix for a spin(qubit) system. For a single spin system, the Hamiltonian is
\be\label{fer9}
\H (t)=\frac{\hbar\om_0}{2}(S_z+Y_z)+\hbar\,g^s(t)\,(S_{-}Y_{+}+S_{+}Y_{-}),
\ee
and Eq.~(\ref{fer8}) reduces to
\be\label{fer10}
_S\la n|\rohat_S (t)|m\ra_S=\sum_{p=0}^1 (-1)^p \tr\Big((S_{+}(t))^{p+m} (S_{-} (t))^{p+n}\rohat(0)\Big).
\ee

\subsection{The Eq.~(\ref{fer10}) fulfills the Lindblad master equation}

For the single spin system, in the interaction picture, the Hamiltonian Eq.~(\ref{fer9}) becomes
\bea\label{ss1}
\H_I (t)=&& \Uhd_0 (t)\hbar\,\big[g^s(t)\,(S_{-}Y_{+}+S_{+}Y_{-})\big]\Uh_0(t),\nn\\
         =&& \hbar\,g^s(t)\,(S_{-}Y_{+}+S_{+}Y_{-}),
\eea
with the free evolution operator
\be\label{ss2}
\Uh_0 (t)=e^{-\frac{i\om_0 t}{2}\,S_z}\ot e^{-\frac{i\om_0 t}{2}\,Y_z}.
\ee
Since $[\H_I (t), \H_I (t')]=0$, the time-evolution operator in the interaction picture is 
\be\label{ss3}
\Uh_I (t)=e^{-ik_s(t)(S_{-}Y_{+}+S_{+}Y_{-})},
\ee
where we have defined
\be\label{ss4}
k_s(t)=\int_0^t dt'\,g^s (t').
\ee
The Schr\"{o}dinger-picture evolution operator is $\Uh (t)=\Uh_0 (t)\Uh_I(t)$. In Eq.~(\ref{fer10}), spin operators are in the Heisenberg picture, so
\bea\label{ss5}
S_{-}(t) &=& (\Uh_0(t)\Uh_I(t))^\dag S_{-} (\Uh_0(t)\Uh_I(t)),\nn\\
         &=& \Uhd_I (t)\big(\Uhd_0 (t)S_{-}\Uh_0 (t)\big)\Uh_I(t),\nn\\
         &=& e^{-i\om_0 t}\,\Uhd_I (t)S_{-}\Uh_I (t),\nn\\
         &=& e^{-i\om_0 t}\,\big[\cos(k_s (t))\, S_{-}+i\sin(k_s(t))\,S_{z}\,Y_{-}\big],\nn\\
\eea
and
\bea\label{ss6}
S_{+}(t) &=& (S_{-}(t))^\dag,\nn\\
         &=& e^{i\om_0 t}\,\big[\cos(k_s (t))\, S_{+}-i\sin(k_s(t))\,S_{z}\ot Y_{+}\big].\nn\\
\eea
Inserting these into Eq.~(\ref{fer10}) gives
\bea\label{ss7}
&& _S\la n|\rohat_S (t)|m\ra_S=\nn\\
&& e^{-i\om_0 t(n-m)}\sum_{p=0}^1 (-1)^p \tr\Big\{\big[\cos(k_s (t)) S_{+}-i\sin(k_s(t)) S_{z}\ot Y_{+}\big]^{p+m}\nn\\
&& \times\,\big[\cos(k_s (t))\, S_{-}+i\sin(k_s(t))\,S_{z}\,Y_{-}\big]^{p+n}\rohat(0)\Big\}.\nn\\
\eea
Using the identities $(S_{\pm} (t))^2=\Uhd (t)S^2_{\pm}\Uh (t)=0$, the summation over $p$ in Eq.~(\ref{ss7}) is restricted to $0\leq s+m\leq 1,$ and $0\leq s+n\leq 1$, which considerably simplifies the calculation. To proceed, let the bath system be initially in the thermal state
\be\label{thermal spin}
\rohat_B (0)=\left(
               \begin{array}{cc}
                 p_1 & 0 \\
                 0 & p_2 \\
               \end{array}
             \right),
\ee
with thermal populations
\begin{equation}\label{pp}
\begin{array}{l@{}}
 p_1=\frac{e^{-\frac{\beta\hbar\om_0}{2}}}{e^{-\frac{\beta\hbar\om_0}{2}}+e^{\frac{\beta\hbar\om_0}{2}}},\\
 p_2=\frac{e^{\frac{\beta\hbar\om_0}{2}}}{e^{-\frac{\beta\hbar\om_0}{2}}+e^{\frac{\beta\hbar\om_0}{2}}},
\end{array}
\end{equation}
and let the main system be initially in an arbitrary state $\rohat_S (0)$. In this case, the explicit components of the reduced density matrix are found to be
\be\label{rosmain}
\rohat_S (t)=\left(
               \begin{array}{cc}
                 \cos^2(k_s (t))\,\rohat_{S, 11} (0)+p_1\,\sin^2(k_s (t)) & e^{-i\om_0 t}\cos(k_s (t))\,\rohat_{S, 12} (0) \\
                 e^{-i\om_0 t}\cos(k_s (t))\,\rohat_{S, 21} (0) & \cos^2(k_s (t))\,\rohat_{S, 22} (0)+p_2\sin^2(k_s (t)) \\
               \end{array}
             \right).
\ee
Taking the time-derivative of both sides of Eq.~(\ref{rosmain}), yields the Lindblad
master equation for a two-level (spin) system interacting with its thermal bath (Appendix \ref{secE})
\bea\label{dotro7}
\dot{\rohat}_S (t) &=& -\frac{i}{\hbar}[\H_S,\rohat_s(t)]+\gamma\,(\bar{n}+1)\,\Big(S_{-}\rohat_S (t)S_{+}-\frac{1}{2}\,\{S_{+}S_{-},\rohat_{S}(t)\}\Big)\nn\\
&& + \gamma\,\bar{n}\,\Big(S_{+}\rohat_S (t)S_{-}-\frac{1}{2}\,\{S_{-}S_{+},\rohat_{S}(t)\}\Big).
\eea

\section{Conclusion}\label{sec13}

In summary, we have developed a direct and computationally efficient method to construct the reduced density matrix of bosonic and fermionic open quantum systems under Markovian dynamics. Our key result is that, by engineering the system-bath coupling to ensure unidirectional energy flow, the resulting reduced dynamics exactly satisfies the Lindblad master equation without requiring the full diagonalization of the Liouvillian. The framework's conceptual simplicity belies its effectiveness, as demonstrated by its accurate reproduction of dissipative behavior across a range of paradigmatic systems.

The method's tractability opens several avenues for future exploration, including the more general time-dependent couplings and extensions to non-Markovian environments. More broadly, this approach establishes a practical bridge between formal open-system theory and large-scale quantum simulations, offering a valuable computational resource for studying quantum thermalization, non-equilibrium transport, and the design of robust quantum information platforms.



\begin{appendices}

\section{Derivation of Eq.~(\ref{iso7})}\label{secA}
Starting from the definition of the reduced density matrix elements, we have
\bea\label{sbos3}
&& _S\la \mbn|\rohat_S (t)|\mbm\ra_S = _S\la \mbf{0}|\frac{(\a_N)^{n_N}}{\sqrt{n_N!}}\cdots\frac{(\a_1)^{n_1}}{\sqrt{n_1!}}\rohat_S (t)\frac{(\ad)^{m_1}}{\sqrt{m_1!}}\cdots\frac{(\ad)^{m_N}}{\sqrt{m_N!}}|\mbf{0}\ra_S,\nn\\
&& = \frac{1}{\sqrt{n_1!m_1!\cdots n_N! m_N!}} _S\la \mbf{0}|\tr_B\Big((\a_N)^{n_N}\cdots (\a_1)^{n_1}\rohat(t)(\ad_1)^{m_1}\cdots(\ad_N)^{m_N}\Big)|\mbf{0}\ra_S,\nn\\
&& = \frac{1}{\sqrt{n_1!m_1!\cdots n_N! m_N!}}\tr_S \Bigg(|\mbf{0}\ra_S\la \mbf{0}|\tr_B\Big((\a_N)^{n_N}\cdots (\a_1)^{n_1}\rohat(t)(\ad_1)^{m_1}\cdots(\ad_N)^{m_N}\Big)\Bigg),\nn\\
\eea
now, by inserting the identity Eq.~(\ref{bos5}) for the vacuum projector $|\mbf{0}\ra_S\la \mbf{0}|$, we obtain 
\bea\label{iso6}
&& _S\la \mbn|\rohat_S (t)|\mbm\ra_S =\frac{1}{\sqrt{n_1!m_1!\cdots n_N! m_N!}}\sum_{s_1,...,s_N=0}^\infty \frac{(-1)^{\sum\limits_{i=1}^N s_i}}{s_1!\cdots s_N!}\nn\\
&& \,\,\,\,\,\,\,\,\,\,\times\,
\tr\Big((\ad_1)^{s_1+m_1}\cdots(\ad_N)^{s_N+m_N}(\a_1)^{s_1+n_1}\cdots(\a_N)^{s_N+n_N}\rohat(t)\Big),
\eea
where the trace is over the total Hilbert space. Using $\rohat (t)=\hat{U} (t)\rohat (0)\hat{U}^\dag (t)$, and the cyclic property of the trace, together with the Heisenberg evolution $\hat{O} (t)=\hat{U}^\dag (t)\hat{O}\hat{U}(t)$, we arrive at Eq.~(\ref{iso7}).

\section{Derivation of Eq.~(\ref{os11})}\label{secB}
By making use of the main formula Eq.~(\ref{iso8}), and the identities
\bea\label{os10}
&& (\ad (t))^{s+m}=(\bar{\mu}(t)\,\ad+\bar{\nu}(t)\,\bd)^{s+m}\nn\\
&&\,\,\,=\sum_{p=0}^{s+m}\binom{s+m}{p}(\bar{\mu}\,\ad)^{s+m-p}(\bar{\nu}\,\bd)^{p},\nn\\
&& (\a (t))^{s+n}=(\mu(t)\,\a+\nu(t)\,\b)^{s+n}\nn\\
&& \,\,\,=\sum_{q=0}^{s+n}\binom{s+n}{q}(\mu\,\a)^{s+n-q}(\nu\,\b)^{q},
\eea
we obtain
\bea
&& _S\la n|\rohat_S (t)|m\ra_S=\nn\\
&& \frac{1}{\sqrt{n!m!}}\sum_{s=0}^\infty \frac{(-1)^s}{s!}\sum_{p=0}^{s+m}\sum_{q=0}^{s+n}\binom{s+m}{p}\binom{s+n}{q}
\underbrace{(\bar{\mu})^{s+m-p}\mu^{s+n-q}(\bar{\nu})^p\nu^q}_{\Gamma(s,n,m,p;t)}\nn\\
&&\,\,\,\,\,\,\times\,\underbrace{\tr_S\Big[(\ad)^{s+m-p}(\a)^{s+n-q}\rohat_S (0)\Big]}_{\chi(s,m,n,p)}\underbrace{\tr_B\Big[(\bd)^p (\b)^q\rohat_B(0)\Big]}_{\delta_{pq}\,p!\,\bar{n}_b^p},\nn\\
&& = \frac{1}{\sqrt{n!m!}}\sum_{s=0}^\infty \frac{(-1)^s}{s!}\sum_{p=0}^{s+m}\binom{s+m}{p}\binom{s+n}{p}\,p!\,\bar{n}_b^p\,\Gamma(s,n,m,p;t)\,\chi(s,m,n,p).\nn\\
\eea
\section{Derivation of Eq.~(\ref{os16})}\label{secC}
\noindent The function $\Gamma_{s,m,n,p}(t) $ for the choice $\cos^2(k_b(t))=e^{-\gamma t}$, becomes
\bea\label{gam}
\Gamma(s,n,m,p;t)&=&(\bar{\mu})^{s+m-p}\mu^{s+n-p}(\bar{\nu})^p\nu^p,\nn\\
                 &=& e^{-t[i\om_0 (n-m)+\gamma (s-p)+\frac{\gamma}{2}(n+m)]}\,(1-e^{-\gamma t})^p.
\eea
Note that, the time-dependence of the density matrix $\rohat_S (t)$ is governed by $\Gamma(s,n,m,p;t)$, therefore, by taking the time derivative of both sides of Eq.~(\ref{os11}), we obtain
\bea\label{os12}
&& _S\la n|\dot{\rohat}_S (t)|m\ra_S=\nn\\
&& \frac{1}{\sqrt{n!m!}}\sum_{s=0}^\infty \frac{(-1)^s}{s!}\sum_{p=0}^{s+m}\sum_{q=0}^{s+n}\binom{s+m}{p}\binom{s+n}{q}\,p!\bar{n}_b^p\,
\,\dot{\Gamma}(s,n,m,p;t)\,\chi(s,m,n,p).\nn\\
\eea
By inserting
\bea\label{os13}
&& \dot{\Gamma}(s,n,m,p;t)=\big(-i\om_0 (n-m)-\gamma(s-p)-\frac{\gamma}{2}(n+m)\big)\,\Gamma(s,n,m,p;t)\nn\\
&& \,\,\,\,\,\,\,\,+\,p\,\big(1-e^{-\gamma t}\big)^{p-1}\,\big(\gamma e^{-\gamma t}\big)\,e^{-t[-i\om_0 (n-m)-\gamma(s-p)-\frac{\gamma}{2}(n+m)]},
\eea
into Eq.~(\ref{os12}) and rearranging the indices $s,n,m$, and $p$, we obtain
\bea\label{os14}
&& _S\la n|\dot{\rohat}_S (t)|m\ra_S=\big(-i\om_0 (n-m)-\gamma(s-p)-\frac{\gamma}{2}(n+m)\big)\,_S\la n|\rohat_S (t)|m\ra_S\nn\\
&& \,\,\,\,\,\,+ \gamma\sqrt{(n+1)(m+1)}\,_S\la n+1|\rohat_S (t)|m+1\ra_S+\gamma\bar{n}_b\sqrt{nm}\,_S\la n-1|\rohat_S (t)|m-1\ra_S\nn\\
&& \,\,\,\,\,\,- \gamma\bar{n}_b (n+m)\,_S\la n|\rohat_S (t)|m\ra_S -\gamma\bar{n}_b\,_S\la n|\rohat_S (t)|m\ra_S\nn\\
&& \,\,\,\,\,\,+ \gamma\bar{n}_b\,\sqrt{(n+1)(m+1)}\,_S\la n+1|\rohat_S (t)|m+1\ra_S.
\eea 
Now, by making use of the identities

\begin{equation}\label{os15}
\begin{array}{l@{}l}
 \sqrt{nm}\,_S\la n-1|\rohat_S (t)|m-1\ra_S =\, _S\la n|\ad\rohat_S (t)\a|m\ra_S,\\
 (n+m)\,_S\la n|\rohat_S (t)|m\ra_S =\, _S\la n|\{\ad\a,\rohat_S (t)\}|m\ra_S,\\
 \sqrt{(n+1)(m+1)}\,_S\la n+1|\rohat_S (t)|m+1\ra_S = \,_S\la n|\a\rohat_S (t)\ad|m\ra_S,\\
 (n-m)\,_S\la n|\rohat_S (t)|m\ra_S= _S\la n|[\ad\a,\rohat_S (t)]|m\ra_S,
\end{array}
\end{equation}
we finally obtain Eq.~(\ref{os16}).
\section{Derivation of Eq.~(\ref{fer8})}\label{secD}
We have
\begin{eqnarray}\label{Dfer4}
&& _S\la n_1,n_2,...,n_N|\rohat_S (t)|m_1,m_2,...,m_N\ra_S =\nn\\
&& _S\la \mbf{0}|(S^N_{-})^{n_N}\cdots(S^2_{-})^{n_2}(S^1_{-})^{n_1}\rohat_S(t)(S^1_{+})^{m_1}(S^2_{+})^{m_2}\cdots (S^N_{+})^{m_N}|\mbf{0}\ra_S =\nn\\
&& \,\tr_S\Big\{|\mbf{0}\ra_S\la\mbf{0}|\tr_B\Big((S^N_{-})^{n_N}\cdots(S^2_{-})^{n_2}(S^1_{-})^{n_1}\rohat(t)(S^1_{+})^{m_1}(S^2_{+})^{m_2}\cdots (S^N_{+})^{m_N}\Big)\Big\}.\nn\\
\end{eqnarray}
For a single spin, we use the identity 
\be\label{fer5}
|0\ra\la 0|=\sum_{p=0}^1 (-1)^p (S_{+})^p(S_{-})^p.
\ee
Extending this to the multi-spin case,
\be\label{fer6}
|\mbf{0}\ra_S\la\mbf{0}|=\sum_{p_1,...,p_N=0}^1(-1)^{\sum\limits_{i=1}^N p_i}(S^1_{+})^{p_1}(S^1_{-})^{p_1}(S^2_{+})^{p_2}(S^2_{-})^{p_2}\cdots(S^N_{+})^{p_N}(S^N_{-})^{p_N},
\ee
and inserting it into Eq.~(\ref{Dfer4}), yields
\bea\label{fer7}
&& _S\la n_1,n_2,...,n_N|\rohat_S (t)|m_1,m_2,...,m_N\ra_S=\nn\\
&& \sum_{p_1,...,p_N=0}^1(-1)^{\sum\limits_{i=1}^N p_i}\,\tr\Big((S^1_{+})^{p_1+m_1}\cdots(S^N_{+})^{p_N+m_N}(S^1_{-})^{p_1+n_1}\cdots(S^N_{-})^{p_N+n_N}\rohat(t)\Big),\nn\\
\eea
now, using $\rohat (t)=\hat{U} (t)\rohat (0)\hat{U}^\dag (t)$, the cyclic property of the trace, and the definition of the Heisenberg picture, we finally obtain Eq.~(\ref{fer8}).
\section{Derivation of Eq.~(\ref{dotro7})}\label{secE}
\noindent
For the choice $\cos^2(k_s (t))=e^{-\gamma_s t}$, we will find the appropriate $\gamma_s$ in the following. We have
\be\label{rosmain2}
\rohat_S (t)=\left(
               \begin{array}{cc}
                e^{-\gamma_s t}\,\rohat_{S, 11} (0)+p_1\,\big(1-e^{-\gamma_s t}\big) & e^{-i\om_0 t}e^{-\frac{\gamma_s t}{2}}\,\rohat_{S, 12} (0) \\
                 e^{i\om_0 t}e^{-\frac{\gamma_s t}{2}}\,\rohat_{S, 21} (0) & e^{-\gamma_s t}\,\rohat_{S, 22} (0)+p_2\,\big(1-e^{-\gamma_s t}\big) \\
               \end{array}
             \right).
\ee
By taking the time-derivative of both sides of Eq.~(\ref{rosmain2}), we obtain
\bea\label{dotro}
\dot{\rohat}_S (t)=\left(
               \begin{array}{cc}
                -\gamma_se^{-\gamma_s t}\,\big(\rohat_{S, 11}(0)-p_1)\big) & -\big(i\om_0 +\frac{\gamma_s}{2}\big)e^{-t\big(i\om_0 +\frac{\gamma_s}{2}\big)}\rohat_{S, 12} (0) \\
                 \big(i\om_0 -\frac{\gamma_s}{2}\big)e^{t\big(i\om_0 -\frac{\gamma_s}{2}\big)}\rohat_{S, 12} (0) & -\gamma_s\,e^{-\gamma_s t}\,\big(\rohat_{S, 22}(0)-p_2)\big) \\
               \end{array}
             \right).
\eea
By making use of the identities
\bea\label{dotro2}
&& -\frac{i}{\hbar}[\H_S,\rohat_s(t)]=-i\om_0\,\left(
                                                 \begin{array}{cc}
                                                   0 & \rohat_{S,12} (t) \\
                                                   \rohat_{S,21} (t) & 0 \\
                                                 \end{array}
                                               \right),\nn\\
&& \{S_{+}S_{-},\rohat_{S}\}=\left(
                               \begin{array}{cc}
                                 2\rohat_{s,11} (t) & \rohat_{S,12} (t) \\
                                 \rohat_{S,21} (t) & 0 \\
                               \end{array}
                             \right),
\eea
Eq.~(\ref{dotro}) can be rewritten as
\bea\label{dotro3}
\dot{\rohat}_S (t) &=& -\frac{i}{\hbar}[\H_S,\rohat_s(t)]-\frac{\gamma_s}{2}\,\{S_{+}S_{-},\rohat_{S}(t)\}+\gamma_s\,\left(
                                                                                                                    \begin{array}{cc}
                                                                                                                      p_1 & 0 \\
                                                                                                                      0 & p_2 \\
                                                                                                                    \end{array}
                                                                                                                  \right)
\nn\\
                   && -\gamma_s\,\left(
                                            \begin{array}{cc}
                                              0 & 0 \\
                                              0 & \underbrace{e^{-\gamma_s t} (\rohat_{S,22}-p_2)+p_2}_{\rohat_{S,22} (t)} \\
                                            \end{array}
                                          \right).
\eea
Now using the identity $\tr_S(\rohat_S)=1$, and
\be\label{dotro4}
S_{-}\rohat_S (t)S_{+}=\left(
                         \begin{array}{cc}
                           0 & 0 \\
                           0 & 1\\
                         \end{array}
                       \right)-\left(
                                 \begin{array}{cc}
                                   0 & 0 \\
                                   0 & \rohat_{S,22} \\
                                 \end{array}
                               \right),
\ee
we can rewrite Eq.~(\ref{dotro3}) as
\bea\label{dotro5}
\dot{\rohat}_S (t) &=& -\frac{i}{\hbar}[\H_S,\rohat_s(t)]-\frac{\gamma_s}{2}\,\{S_{+}S_{-},\rohat_{S}(t)\}+\gamma_s\,\left(
                                                                                                                    \begin{array}{cc}
                                                                                                                      p_1 & 0 \\
                                                                                                                      0 & -p_1 \\
                                                                                                                    \end{array}
                                                                                                                  \right)
\nn\\
                   && + \gamma_s\,S_{-}\rohat_S (t)S_{+}.
\eea
Finally, if we set $\gamma_s=\gamma(2\bar{n}+1)$, where $\bar{n}=(e^{\beta\hbar\om_0}-1)^{-1}$, and make use of the identity
\be\label{dotro6}
S_{+}\rohat_S (t)\rohat_{-}-\frac{1}{2}\{S_{-}S_{+},\rohat_S (t)\}=\left(
                                                                     \begin{array}{cc}
                                                                       \rohat_{S,22} & -\frac{1}{2}\,\rohat_{S,12} \\
                                                                       -\frac{1}{2}\,\rohat_{S,21} & -\rohat_{S,22} \\
                                                                     \end{array}
                                                                   \right),
\ee
we finally obtain Eq.~(\ref{dotro7}).

\end{appendices}


\begin{thebibliography}{}
\bibitem{Breuer2007} Breuer, H. P. \& Petruccione F. \emph{The Theory of Open Quantum Systems} (Oxford University Press, Oxford, 2007).
\bibitem{Weiss2012} Weiss, U. \emph{Quantum Dissipative Systems}, 4th ed. (World Scientific, Singapore, 2012).
\bibitem{Rivas2012} Rivas, A. \& Huelga, S. F. \emph{Open Quantum Systems: An Introduction} (Springer, Berlin, 2012).
\bibitem{Nakajima1958} Nakajima, S. \emph{On Quantum Theory of Transport Phenomena: Steady Diffusion}. Prog. Theor. Phys. \textbf{20}, 948 (1958).
\bibitem{Zwanzig1960} Zwanzig, R. J. \emph{Ensemble Method in the Theory of Irreversibility}. Chem. Phys. \textbf{33}, 1338 (1960).
\bibitem{Fogedby2022} Fogedby, H. C. \emph{Field-theoretical approach to open quantum systems and the Lindblad equation}. Phys. Rev. A \textbf{106}, 022205 (2022).
\bibitem{Lidar2001}  Lidar, D. A., Bihary, Z., \& Whaley, K. B. \emph{From completely positive maps to the quantum Markovian semigroup master equation}. Chem. Phys. \textbf{268}, 35 (2001).
\bibitem{Lindblad1976} Lindblad, G. \emph{On the generators of quantum dynamical semigroups}. Commun. Math. Phys. \textbf{48}, 119 (1976).
\bibitem{Gorini1976} Gorini, V., Kossakowski, A., \& Sudarshan E. C. G. J. \emph{Completely positive dynamical semigroups of N-level systems}.
 Math. Phys. \textbf{17}, 821 (1976).
\bibitem{Manzano2020} Manzano, D. \emph{A short introduction to the Lindblad master equation}. AIP Advances \textbf{10}, 025106 (2020).
\bibitem{Thompson2023} Thompson, F. \& Kamenev, A. \emph{Field theory of many-body Lindbladian dynamics}. Ann. Phys. \textbf{455}, 169385 (2023).
\bibitem{Barthel2022} Barthel, T. \& Y. Zhang, Y. J. \emph{Solving quasi-free and quadratic Lindblad master equations for open fermionic and bosonic systems}. Stat. Mech. \textbf{11}, 113101 (2022).
\bibitem{Nielsen2010} Nielsen, M. A. \& Chuang, I. L. \emph{Quantum Computation and Quantum Information} (Cambridge University Press, Cambridge, 2010).
\bibitem{Gardiner2004} Gardiner, C. W. \& Zoller, P. \emph{Quantum Noise}, 3rd ed. (Springer, Berlin, 2004).
\bibitem{Walls1994} Walls, D. F. \& Milburn, G. J. \emph{Quantum Optics}, 3rd ed., (Springer, Berlin, 2025).
\bibitem{Carmichael1993} Carmichael H. J. \emph{An Open Systems Approach to Quantum Optics} (Springer, Berlin, 1993).
\bibitem{Sieberer2016} Sieberer, L. M., Buchhold, M. \& Diehl, S. \emph{Keldysh field theory for driven open quantum systems}. Rep. Prog. Phys. \textbf{79}, 096001 (2016).
\bibitem{Kamenev2011} Kamenev, A. \emph{ Field Theory of Non-Equilibrium Systems} (Cambridge University Press, Cambridge, 2011).
\bibitem{Prosen2008} Prosen, T. \emph{Third quantization: a general method to solve master equations for quadratic open Fermi systems} New J. Phys. \textbf{10}, 043026 (2008).
\bibitem{Prosen2010} Prosen, T. \& Seligman, T. H. \emph{Quantization over boson operator spaces} J. Phys. A \textbf{43}, 392004 (2010).
\bibitem{Adeso2007} Adesso, G. \& Illuminati, F., \emph{Entanglement in continuous-variable systems: recent advances and current perspectives} Journal of Physics A: Mathematical and Theoretical, \textbf{40}, 28, 2007
\bibitem {Weedbrook2012} Weedbrook, C. et al. \emph{Gaussian quantum information} Rev. Mod. Phys. \textbf{84}, 621 (2012).
\bibitem{Kheir1-2025} Kheirandish, F. et al. \emph{A novel scheme for modelling dissipation (gain) and thermalization
in open quantum systems} Phys. Scr. \textbf{100} 015110 (2025).
\bibitem{Kheir2-2025} Cheraghpour, N. \& Kheirandish, F. \emph{Quantum dynamics of a bosonic mode and a two-level system interacting with several reservoirs} Laser Phys. \textbf{35} 055204 (2025).
\bibitem{Louisel1975} Louisell, W. H. \emph{Quantum Statistical Properties of Radiation} (Wiley, New York, 1975).
\bibitem{Knight2005} Gerry, C. C. \& Knight P. L.\emph{ Introductory quantum optics} (Cambridge University Press, 2005).
\end{thebibliography}
\section*{Acknowledgements}
\noindent
This work has been supported by the University of Kurdistan. The authors thank Vice Chancellorship of Research and Technology, University of Kurdistan.

\section*{Author contributions}
\noindent
F.K. made the main calculations, F.K., A.D. discussed the results, Z.I., F.B. wrote the paper.

\section*{Additional information}
\noindent
\textbf{Competing financial interests:} The authors declare no competing financial interests.

\end{document}